\documentclass[twocolumn,showpacs]{revtex4-1}
\usepackage{subcaption} \usepackage{graphicx} \usepackage{amsmath}
\graphicspath{{./fig/}}

\begin{document} 

\title[Beyond Nuclear Pasta]{Beyond Nuclear Pasta: Phase Transitions and Neutrino Opacity of Non-Traditional Pasta}

\author{P. N. Alcain, P. A. Gim\'enez Molinelli and C. O. Dorso}

\affiliation{Departamento de F\'isica, FCEyN, UBA and IFIBA, Conicet, 
Pabell\'on 1, Ciudad Universitaria, 1428 Buenos Aires, Argentina} 
\affiliation{IFIBA-CONICET}

\date{\today} 

\begin{abstract}
In this work, we focus on different length scales within the 
dynamics of nucleons in conditions according to the neutron star
crust, with a semiclassical molecular dynamics model, studying
isospin symmetric matter at subsaturation densities. While varying
the temperature, we find that a solid-liquid phase transition 
exists, that can be also characterized with a morphology transition.
For higher temperatures, above this phase transition, we study the
neutrino opacity, and find that in the liquid phase, the scattering
of low momenta neutrinos remain high, even though the morphology
of the structures differ significatively from those of the traditional
nuclear pasta.
\end{abstract}

\maketitle

\section{Introduction}
  \label{intro}
  \makeatletter{}A neutron star has a radius of approximately 10 km and a mass of a few
solar masses. According to current models~\cite{page04}, the structure
neutron star can be roughly described as composed in two parts: the
\emph{crust}, of 1.5 km thick and a density of up to half the normal
nuclear density $\rho_0$, with the structures known as \emph{nuclear
  pasta}~\cite{20}; and the \emph{core}, where the structure is still
unknown and remains highly speculative~\cite{woosley05}. Most neutron
stars are remanents of core collapse supernovae.  This kind of
supernovae happen when the hot and dense Fe core of a dying massive
star (known as proto-neutron star) collapses.  During the collapse,
several nuclear processes take place in the inner core of the star
--- electron capture, photodisintegration, URCA, etc.  These processes,
apart from increasing the overall neutron number of the system,
produce a large amount of neutrinos which flow outwards.  The
interaction between the neutrinos streaming from the core of the
proto-neutron star and its outer layers play an important role in
reversing the collapse that causes the supernova.

On the other hand, neutron stars are born hot, but cool down by
means of neutrino emission. Therefore, the interaction between the
neutrinos and neutron star matter is key to comprehend two aspects of
a neutron star history: its genesis and its thermal evolution.

The neutron stars' crust is composed of neutrons and protons embedded 
in a degenerate electron gas.
Protons and neutrons in the crust are supposedly arranged in
structures that differ substantially from the ``normal'' nuclei ---
the non-homogeneous phases collectively known as \emph{nuclear pasta}. 
The structure of this nuclear pasta is related to the neutrino 
opacity of neutron stars' crust, neutron star quakes, and pulsar 
glitches. Specifically, the neutron star quakes and pulsar glitches 
are related to the mechanical properties of the crust
matter~\cite{32}, while the neutrino opacity is enhanced by coherent 
scattering. This enhancement of the crust's opacity is related to 
the static structure factor of nuclear pasta~\cite{horo_lambda}:
\begin{equation*}
 \frac{\text{d}\sigma}{\text{d}\Omega} =  
 \left( \frac{\text{d}\sigma}{\text{d}\Omega} \right)_{\text{uniform}}
 \times S(q).
\end{equation*}

And since neutron stars cooling is associated with neutrino 
emission from the core, the interaction between the neutrinos and 
the particular structure of the crust would dramatically affect the 
thermal history of young neutron stars.

Several models have been developed to study nuclear pasta, and they
have shown that these structures arise due to the interplay between
nuclear and Coulomb forces in an infinite medium. Nevertheless, the
dependence of the thermodynamic observables has not been studied in
depth.

The original works of Ravenhall \emph{et al.}~\cite{20} and Hashimoto
\emph{et al.}~\cite{21} used a compressible liquid drop model, and
proposed the now known as \emph{pasta phases} --\emph{lasagna},
\emph{spaghetti} and \emph{gnocchi}--. Further works~\cite{22, 23, 24,
  25, 28, 30} analyzed only these phases, with mean field
models. Other authors~\cite{26, 27, 29, horo_gr} used dynamical
models, but focusing mainly on the phases already mentioned. The work
by Nakazato \emph{et al.}~\cite{nakazato09}, inspired by polymer
systems, found also gyroid and double-diamond structures, with a
compressible liquid drop model. Dorso \emph{et al.}~\cite{Dor12}
arrived to pasta phases different from those already mentioned with
molecular dynamics, studying mostly its characterization at very low
temperatures.

In this work, we study the neutron stars with molecular dynamics,
using the CMD model~\cite{pandha}. We use morphologic and thermodynamic 
tools to characterize symmetric neutron star matter at different 
temperatures, focusing on the static structure factor. We characterize 
the morphology of the emerging structures with the pair distribution 
function and the \emph{Minkowski functionals}~\cite{29}, a complete set 
of morphological measures. The possible existence of a solid-liquid 
phase transition is explored via the Lindemann coefficient~\cite{lind}.

With these tools at very low temperatures, we found a phase transition 
that is both morphologic and thermodynamic: a discontinuity in the 
Minkowski functionals shows the morphologic transition, while a coincident 
discontinuity in both energy and the Lindemann coefficient show a 
thermodynamic solid-liquid transition. At temperatures slightly higher, 
we found stable structures that do not belong to usual pasta menu. 
Moreover, these unusual shapes absorb neutrinos more efficiently than
the typical nuclear pasta.

This work is structured as follows.
In section~\ref{cmd} we review the interaction model, along 
with the tools used to analyze the results we obtained.
In section~\ref{phase_transition} we present and characterize the 
solid-liquid phase transition in nuclear pasta. Then, in 
sections~\ref{very_long} and~\ref{transport} we study the very-long 
range order of the pasta phases, focusing on its influence on the 
opacity of the crust to low-momentum neutrinos. We discuss 
the appearance of some unusual pasta shapes, and how these newfound 
structures may enhance the opacity of the crust more than traditional 
\emph{pasta} in section~\ref{unusual_pasta}.
Finally, in section~\ref{discussion} we draw conclusions about the 
results presented.

\section{Classical Molecular Dynamics Model}
  \label{cmd}
  \makeatletter{}Of the many models used to study nuclear pasta, the advantage of
classical or semiclassical models is the accessibility to position and
momentum of all particles at all times. This allows the study of the
structure of the nuclear medium from a particle-wise point of
view. Many models exist with this goal, like simple-semiclassical
potential~\cite{horo_gr}, quantum molecular dynamics~\cite{maruyama98}
and classical molecular dynamics~\cite{pandha}. In these models the
Pauli repulsion between nucleons of equal isospin is either hard-coded
in the interaction or as a separate term~\cite{dorso88}.

In this work, we model the interaction between nucleons with a
classical molecular dynamics (CMD) model to study nuclear
reactions. Dorso \emph{et al.} provided justification for its use in
the stellar crust environment~\cite{Dor12}.

The classical molecular dynamics model CMD, as introduced in
Ref.~\cite{14a}, has been successfully used in heavy-ion reaction
studies to: help understand experimental data~\cite{Che02}; identify
phase-transition signals and other critical
phenomena~\cite{16a,Bar07,CritExp-1,CritExp-2}; and explore the
caloric curve~\cite{EntropyCalCur} and
isoscaling~\cite{8a,Dor11}. CMD uses two two-body potentials to
describe the interaction of nucleons, which are a combination of
Yukawa potentials:
\begin{align*}
  V_{np}(r) &=v_{r}\exp(-\mu_{r}r)/{r}-v_{a}\exp(-\mu_{a}r)/{r}\\ 
  V_{nn}(r) &=v_{0}\exp(-\mu_{0}r)/{r}
  \label{2BP}
\end{align*}
where $V_{np}$ is the potential between a neutron and a proton, and
$V_{nn}$ is the repulsive interaction between either $nn$ or $pp$. The
cutoff radius is $r_c=5.4\,\text{fm}$ and for $r>r_c$ both potentials
are set to zero. The Yukawa parameters $\mu_r$, $\mu_a$ and $\mu_0$
were determined to yield an equilibrium density of $\rho_0=0.16
\,\text{fm}^{-3}$, a binding energy $E(\rho_0)=16
\,\text{MeV/nucleon}$ and a compressibility of
$250\,\text{MeV}$~\cite{pandha}.

To simulate an infinite medium, we used CMD under periodic boundary
conditions, symmetric in isospin (i.e. with $x=Z/A=0.5$, $2500$
protons and $2500$ neutrons) in cubical boxes with sizes adjusted to
have densities between $\rho=0.03 \,\text{fm}^{-3} \le \rho \le
0.13\,\text{fm}^{-3}$. Although in the actual neutron stars the
proton fraction is low ($x<0.5$), we chose to work with symmetric
matter because that way we could study the neutron stars without
having a symmetry term in the energy.

\subsection{Coulomb interaction in the Model}
  \label{coulomb}
  \makeatletter{}Since a neutralizing electron gas embeds the nucleons in the neutron
star crust, the Coulomb forces between protons are screened. One of
the many ways to model this screening effect is the Thomas--Fermi
approximation, used with various nuclear
models~\cite{26,Dor12,horo_lambda}. According to this approximation,
protons interact via a Yukawa-like potential, with a screening length
$\lambda$:
\begin{equation*}
 V_{TF}(r) = q^2\frac{e^{-r/\lambda}}{r}.
\end{equation*}

Theoretical estimations for the screening length $\lambda$ are
$\lambda\sim100\,\text{fm}$~\cite{fetter}, but we set the screening
length to $\lambda=20\,\text{fm}$. This choice was based on previous
studies~\cite{nos-lambda}, where we have shown that this value is
enough to adequately reproduce the expected length scale of density
fluctuations for this model. 

\subsection{Simulation procedure}
  \label{procedure}
  \makeatletter{}The trajectories of the nucleons are then governed by the
Pandharipande and the screened Coulomb potentials. The nuclear system
is cooled from $T=0.8\,\text{MeV}$ to $T=0.2\,\text{MeV}$ using
isothermal molecular dynamics with the Nos\'e--Hoover thermostat
procedure~\cite{nose-hoover}, in the LAMMPS
package~\cite{LAMMPS}. Systems are cooled in small temperature steps
($\Delta T=0.01$), decreasing the temperature once both the energy and
the temperature, as well as their fluctuations, are stable.

\subsection{Other tools}
  \label{other_tools}
  \makeatletter{}\subsubsection{Minkowski functionals}

The Minkowski functionals~\cite{michielsen} are a set of independent
functionals that satisfy three properties: motion invariance,
additivity and continuity. The morphological and topological properties 
of any given closed and oriented surface can be completely characterized 
by these Mikowski functionals~\cite{nos-lambda, wata-2003}. According to 
Hadwiger's theorem~\cite{klain}, for a body in a \emph{d}-dimensional 
Euclidean space, there are $d+1$ Minkowski functionals. For three 
dimensional bodies, these four functionals are: volume $V$, surface area 
$S$, Euler characteristic $\chi$ and integral mean curvature $B$.

While the volume and surface area have an easy intuitive
interpretation, we will explain further the other two functionals. The
Euler characteristic $\chi$ is a topological measure that can be
interpreted as:
\begin{equation*}
  \chi = \text{(isolated regions) + (cavities) - (tunnels)}
\end{equation*}.
The integral mean curvature, known as \emph{mean breadth}, is a
measure of the typical width of the body.

Using these functionals, we proposed a classification for the structures 
obtained in simulations of Neutron Star Matter within the CMD framework 
that is summarized in table~\ref{tab:minko}~\cite{Dor12}.

\begin{table}[ht]
  \centering
  \caption{Classification of NSM shapes based on Minkowski functionals}
  \begin{tabular}{c|| c | c | c} \hline & B $<0$ & B $\sim 0$ & B $>0$ \\
  \hline\hline $\chi$ $>0$ & Anti-Gnocchi & & Gnocchi \\ $\chi$ $\sim0$ &
  Anti-Spaghetti & Lasagna & Spaghetti \\ $\chi$ $<0$ & Anti-Jungle Gym &
  & Jungle Gym \\ [1ex] \hline
  \end{tabular}
\label{tab:minko}
\end{table}

\subsubsection{Lindemann coefficient}

The Lindemann coefficient~\cite{lind} is based on the idea of particle
``disorder'', used mostly to study solid-liquid transitions in
infinite systems like crystals~\cite{bilgram87} and is defined from
the standard deviation of the positions of the particles $\Delta
r_i^2$ as follows :

\begin{equation*}
\Delta_L = \frac{\sqrt{\sum_i\langle\Delta r_i^2/N\rangle}}{a}
\end{equation*}

\noindent where $a$ is the lattice constant of the crystal and $N$ is the total
number of particles. In our case, we set $a=(V/N)^{1/3}$, the typical
length of each particle.

\section{Results and discussion}
  \label{results}
  \makeatletter{}In what follows, we present the results obtained cooling
down a system of approximately 5000 particles.

\subsection{Phase transition}
  \label{phase_transition}

\subsubsection{Thermodinamical Phase Transition}
  \label{thermo}
\begin{figure}[h!]  \centering
\includegraphics[width=\columnwidth]{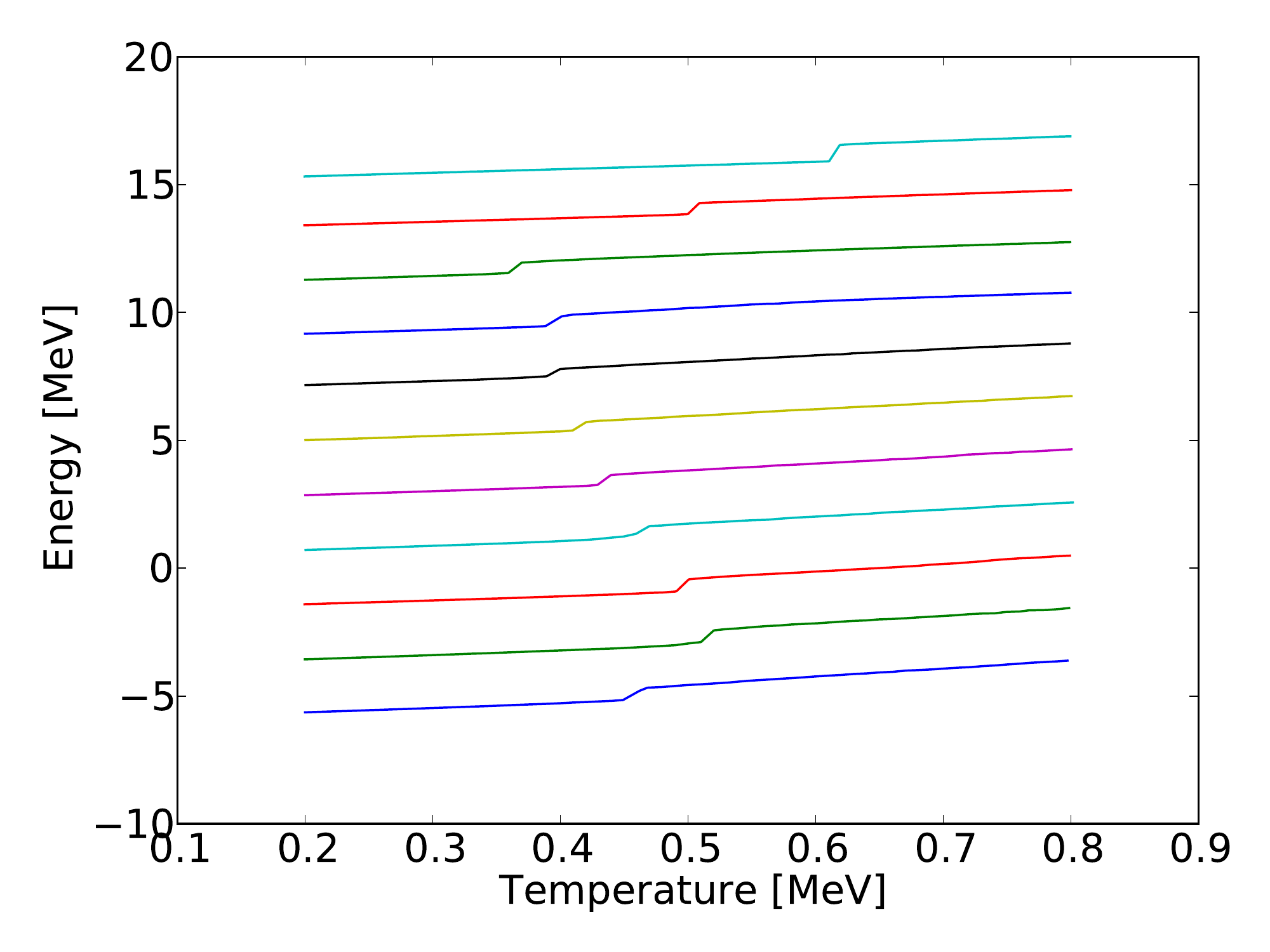}
\caption{Energy as a function of temperature for different
  densities. We see that there is a discontinuity in the range of
  $T_l=0.35\,\text{MeV}$ to $T_h=0.65\,\text{MeV}$, depending on the
  density, a signal of a first-order phase transition. In the figure,
  densities range from $\rho=0.03\,\text{fm}^{-3}$ and
  $\rho=0.13\,\text{fm}^{-3}$, increasing
  $\Delta\rho=0.01\,\text{fm}^{-3}$ upwards.}
\label{fig:energy}
\end{figure}

\begin{figure}[h!]  \centering
\includegraphics[width=\columnwidth]{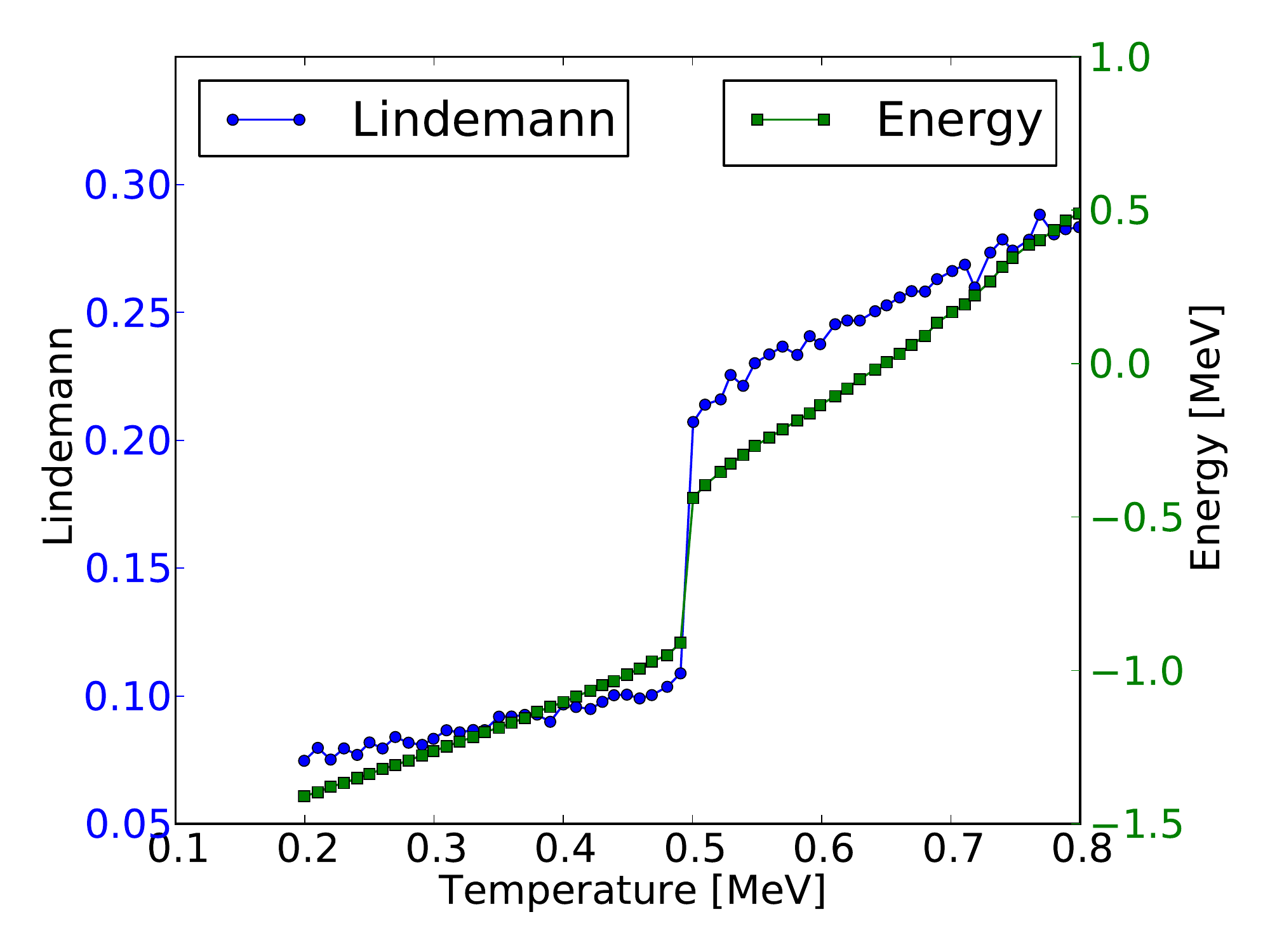}
\caption{Lindemann coefficient and energy as a function of temperature
  for a chosen density, $\rho=0.05\,\text{fm}^{-3}$. The sudden change
  in their value is a signal of a solid-liquid phase transition. We
  can see that both discontinuities are at the same temperature}
\label{fig:lind-energ}
\end{figure}

  \makeatletter{}Figure~\ref{fig:energy} presents the energy as a function of
temperature (caloric curve), for several densities. Each of these
densities exhibit a discontinuity in the energy at certain
temperatures --- a signal of a first order phase transition.  This
transition can be confirmed and further characterized as a
solid-liquid phase transition by looking at the Lindemann
coefficient. The Lindemann coefficient for $\rho=0.05\,\text{fm}^{-3}$
as a function of temperature, can be seen in
figure~\ref{fig:lind-energ}, along the energy. This figure shows that
the discontinuities in Lindemann coefficient and in energy are at the
same temperature. These two factors are, effectively, the signature of
a solid-liquid phase transition.

\begin{figure*}[floatfix]  
  \centering
  \begin{subfigure}[h!]{0.95\columnwidth}
    \includegraphics[width=\columnwidth]{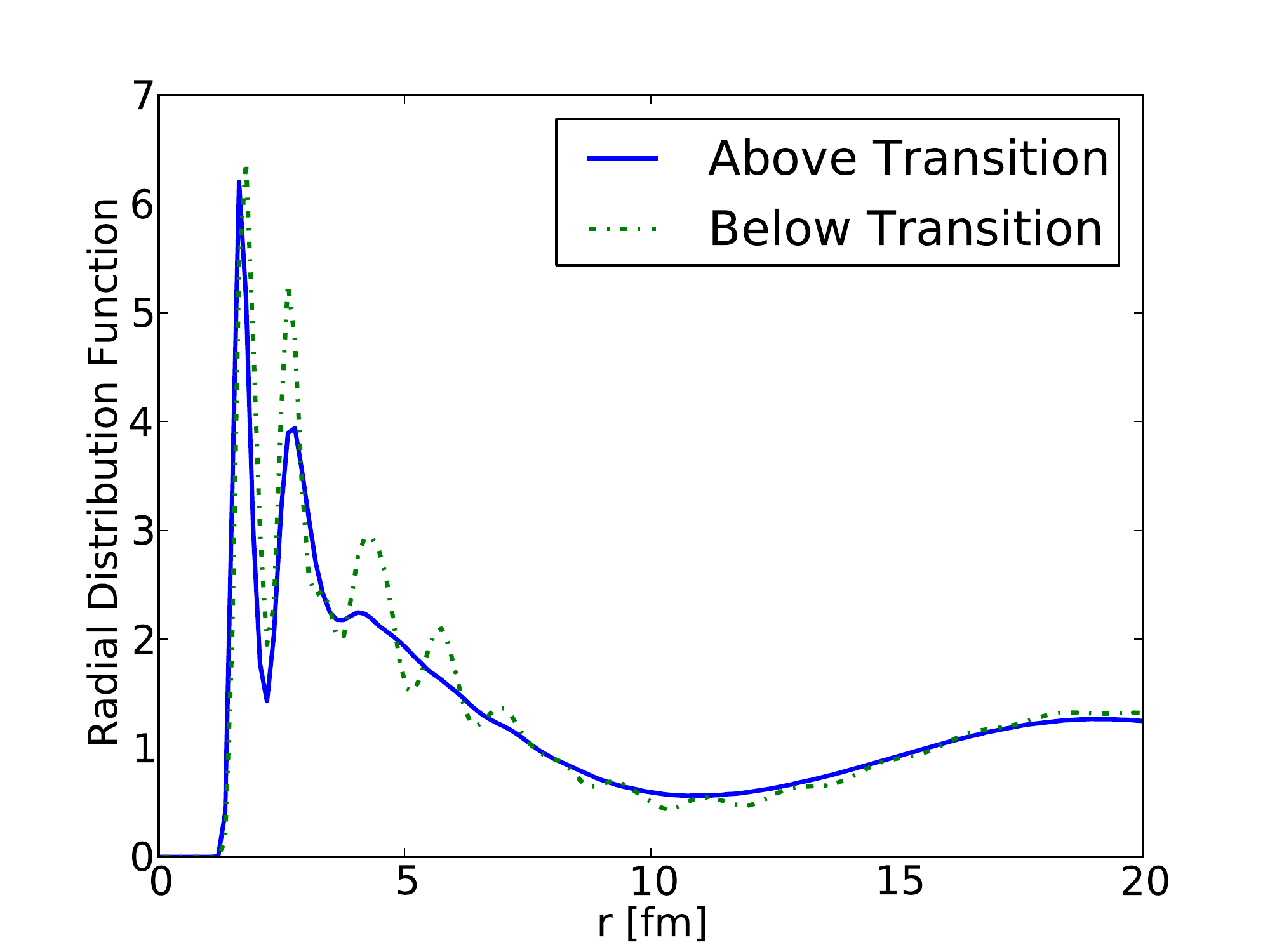}
    \caption*{Radial distribution function for $\rho=0.03\,\text{fm}^{-3}$}
  \end{subfigure}
  \begin{subfigure}[h!]{0.70\columnwidth}
    \includegraphics[width=\columnwidth]{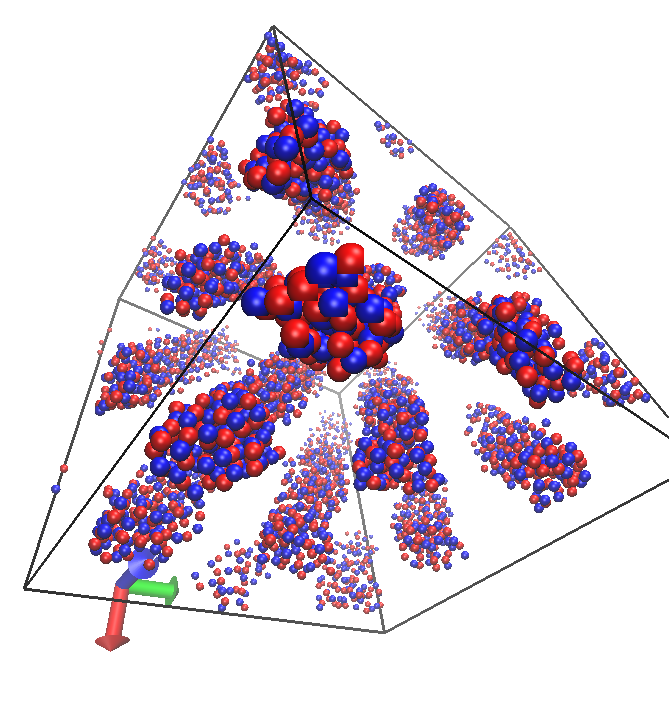}
    \caption*{Snapshot of the system in the liquid phase for $\rho=0.03\,\text{fm}^{-3}$}
  \end{subfigure}
  \begin{subfigure}[h!]{0.95\columnwidth}
    \includegraphics[width=\columnwidth]{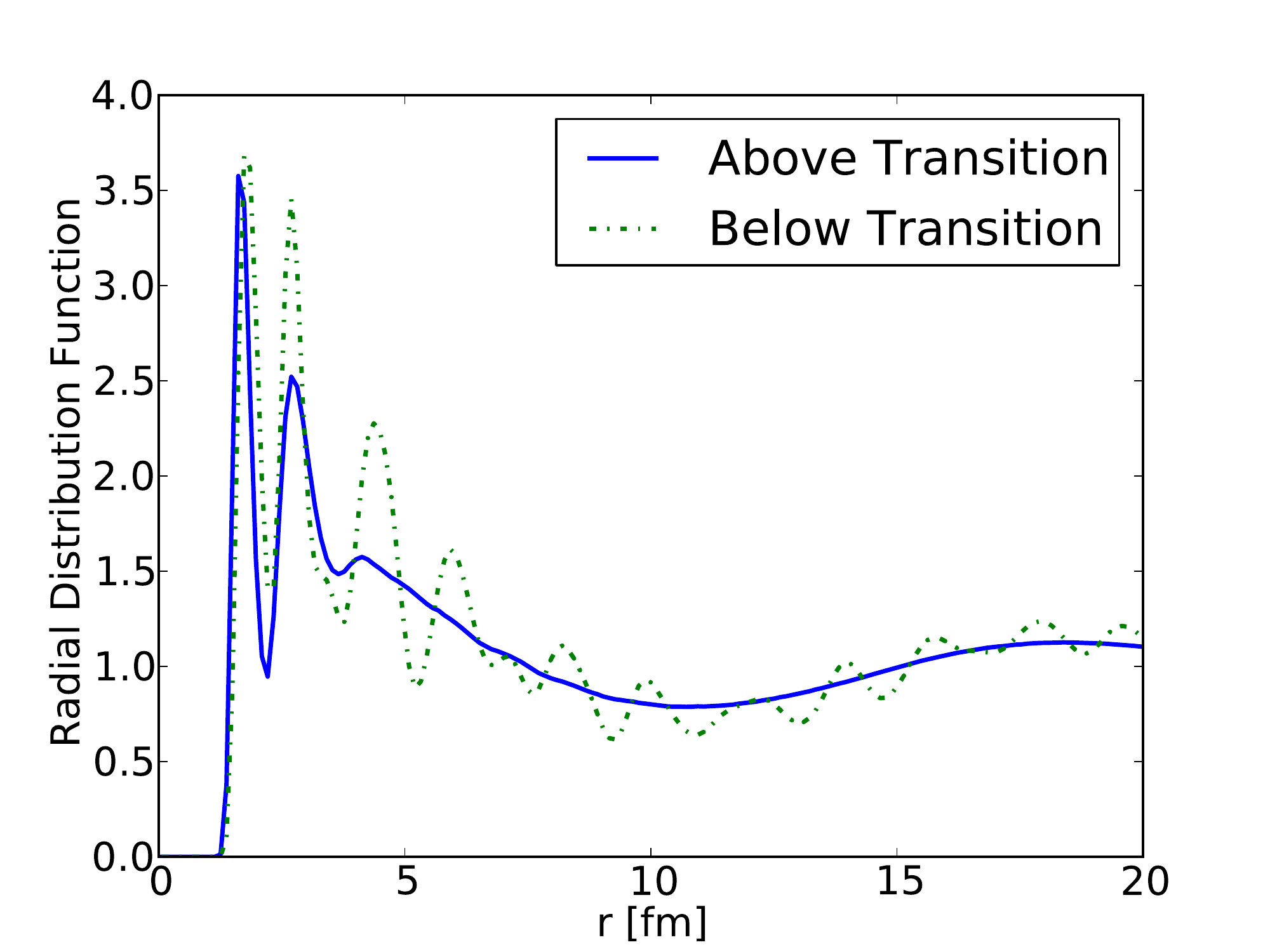}
    \caption*{Radial distribution function for $\rho=0.05\,\text{fm}^{-3}$}
  \end{subfigure}
  \begin{subfigure}[h!]{0.70\columnwidth}
    \includegraphics[width=\columnwidth]{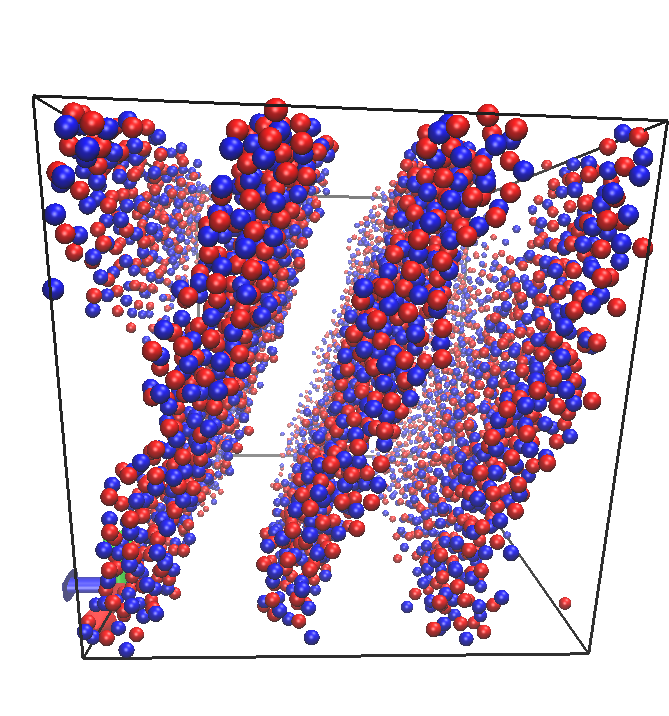}
    \caption*{Snapshot of the system in the liquid phase for $\rho=0.05\,\text{fm}^{-3}$}
  \end{subfigure}
  \begin{subfigure}[h!]{0.95\columnwidth}
    \includegraphics[width=\columnwidth]{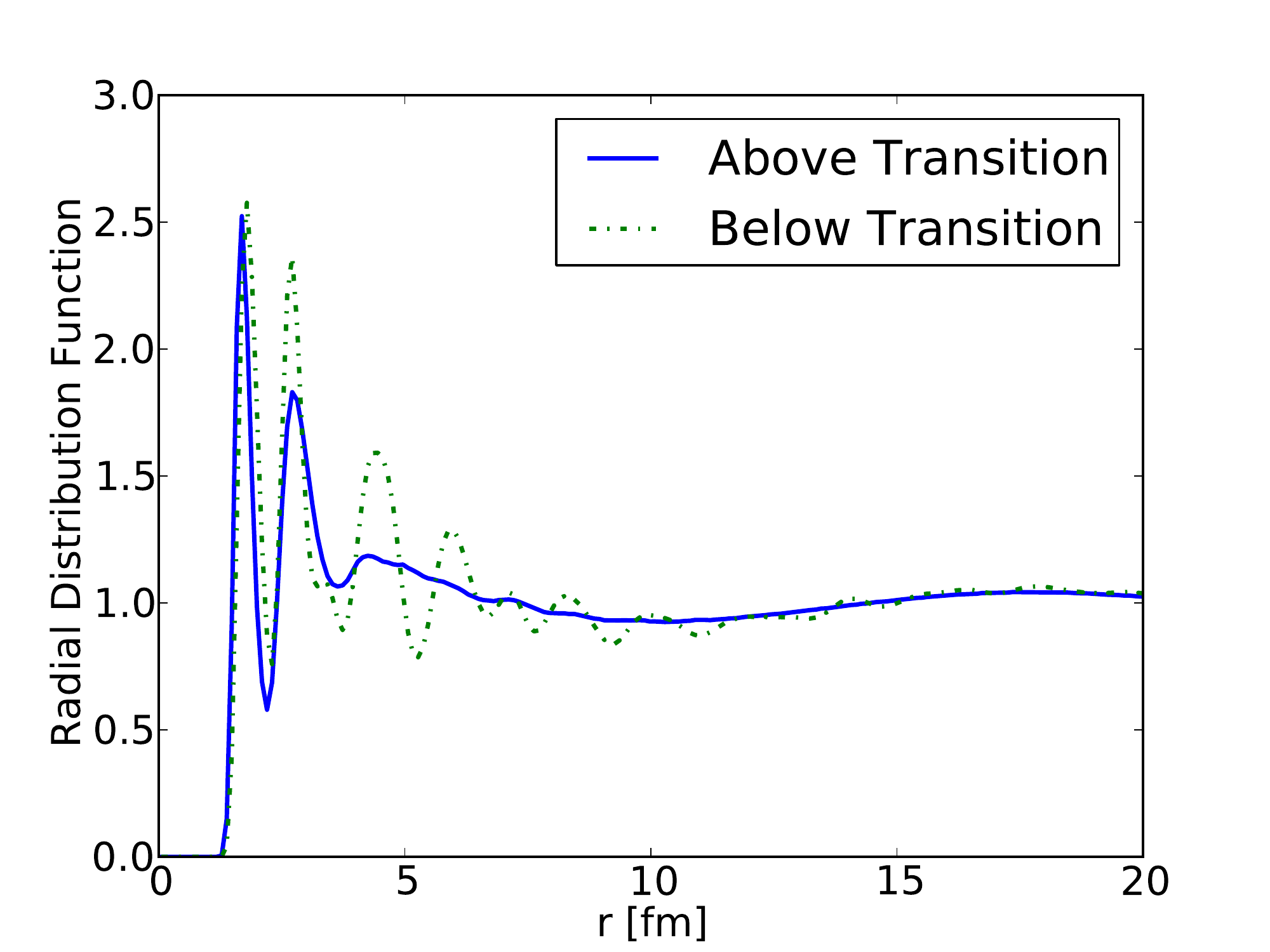}
    \caption*{Radial distribution function for $\rho=0.08\,\text{fm}^{-3}$}
  \end{subfigure}
  \begin{subfigure}[h!]{0.70\columnwidth}
    \includegraphics[width=\columnwidth]{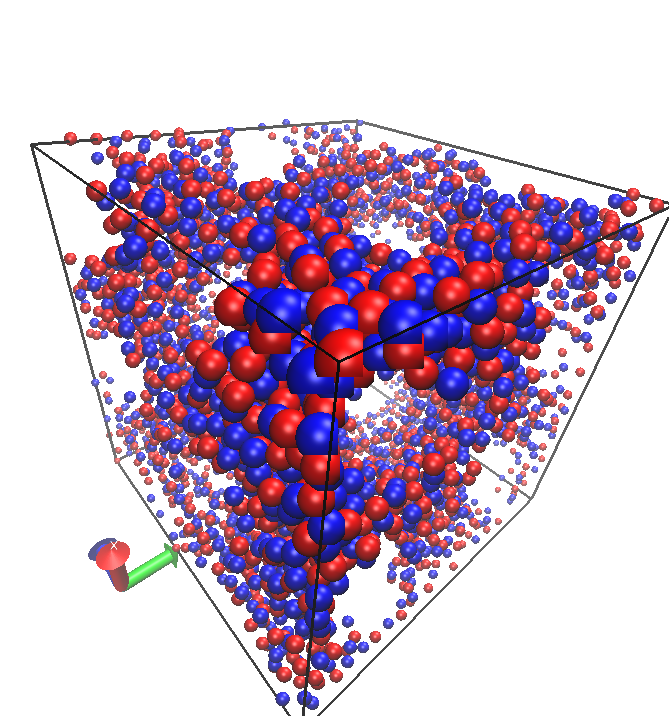}
    \caption*{Snapshot of the system in the liquid phase for $\rho=0.08\,\text{fm}^{-3}$}
  \end{subfigure}
  \caption{Radial distribution function for different densities, both
    below and above the transition temperature, and snapshots of the
    system in the liquid phase. Although the first peaks of the
    distribution are in the same position for both temperatures, the
    following peaks, which exhibit a long-range order typical of
    solids, are only present below the transition temperature.}
  \label{fig:rdf}
\end{figure*}

In figure~\ref{fig:rdf} we show the radial distribution function for
three different densities: $\rho=0.03\,\text{fm}^{-3}$
(\emph{spaghetti}), $\rho=0.05\,\text{fm}^{-3}$ (\emph{lasagna}) and
$\rho=0.08\,\text{fm}^{-3}$ (tunnels), just above and below the
transition temperature, as well as a snapshot of the system at the
high temperature phase. Since the first peaks (corresponding to the
nearest neighbors) are at the same position regardless of the
temperature, we conclude that the short range order is present both
above and below the transition. However, the peaks for third and
higher order neighbors, distinctive of solid phases, disappear as the
temperature is increased through the transition. The very-long range 
orther also survives the transition, as discussed further in section 
\ref{very_long}

\subsubsection{Morphological Phase Transition}
  \label{morpho}
\begin{figure}  \centering \includegraphics[width=\columnwidth]{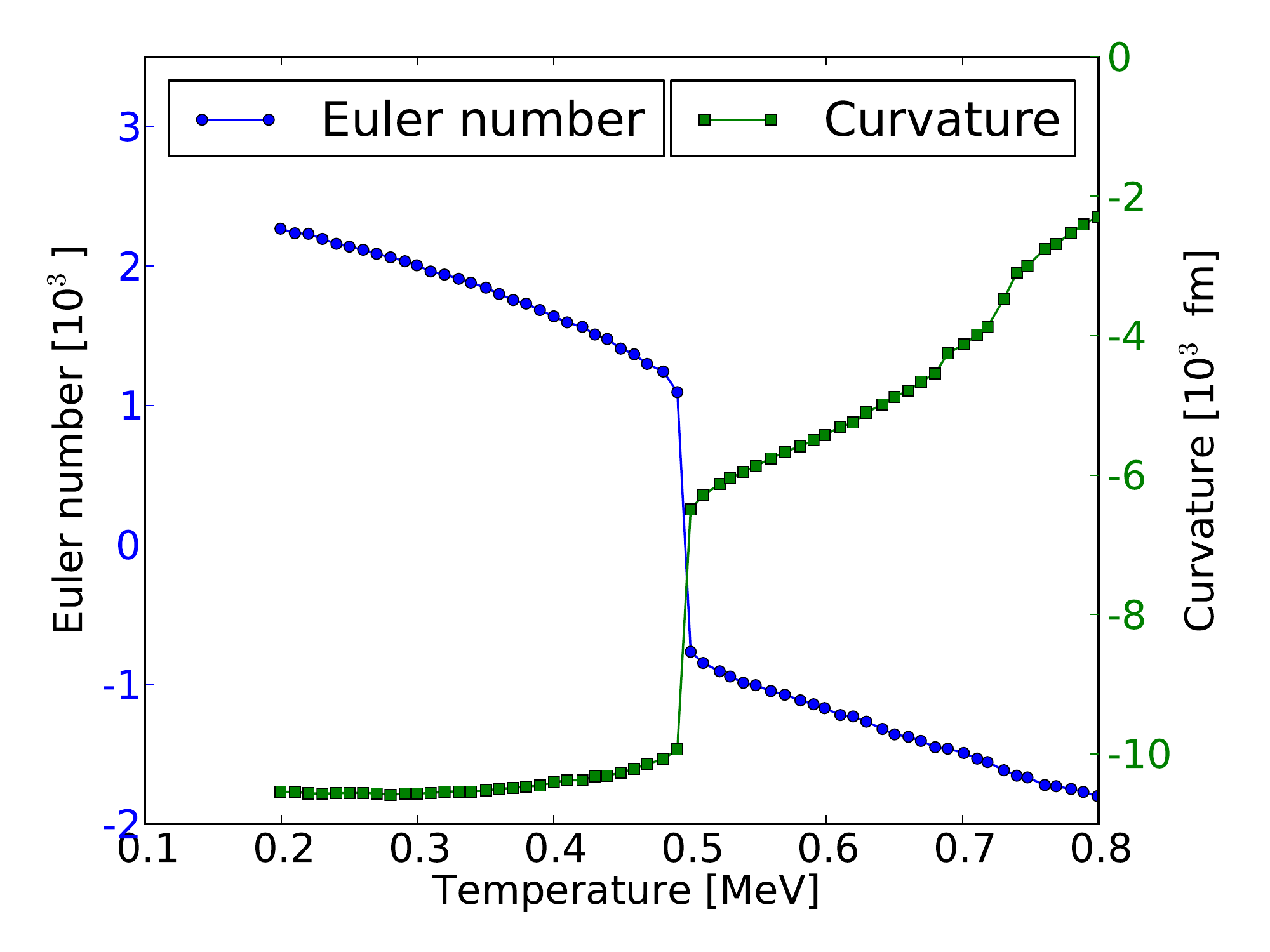}
  \caption{Euler number and mean breadth for $\rho=0.05\,\text{fm}^{-3}$.
    We observe a sharp transition for both Minkowski functionals.}
  \label{fig:euler-curv}
\end{figure}

  \makeatletter{}When we look at the Minkowski functionals, particularly the Euler
characteristic and the mean breadth, we can see that there is again a
``critical'' temperature at which both the Euler characeristic and the
mean breadth show a sharp transition. We show, as an example, these
magnitudes as a function of temperature for density
$\rho=0.05\,\text{fm}^{-3}$ in figure~\ref{fig:euler-curv}. As this
transition is signaled by morphological observables, we conclude that
this transition is morphological.

\begin{figure}[floatfix]  \centering
  \includegraphics[width=\columnwidth]{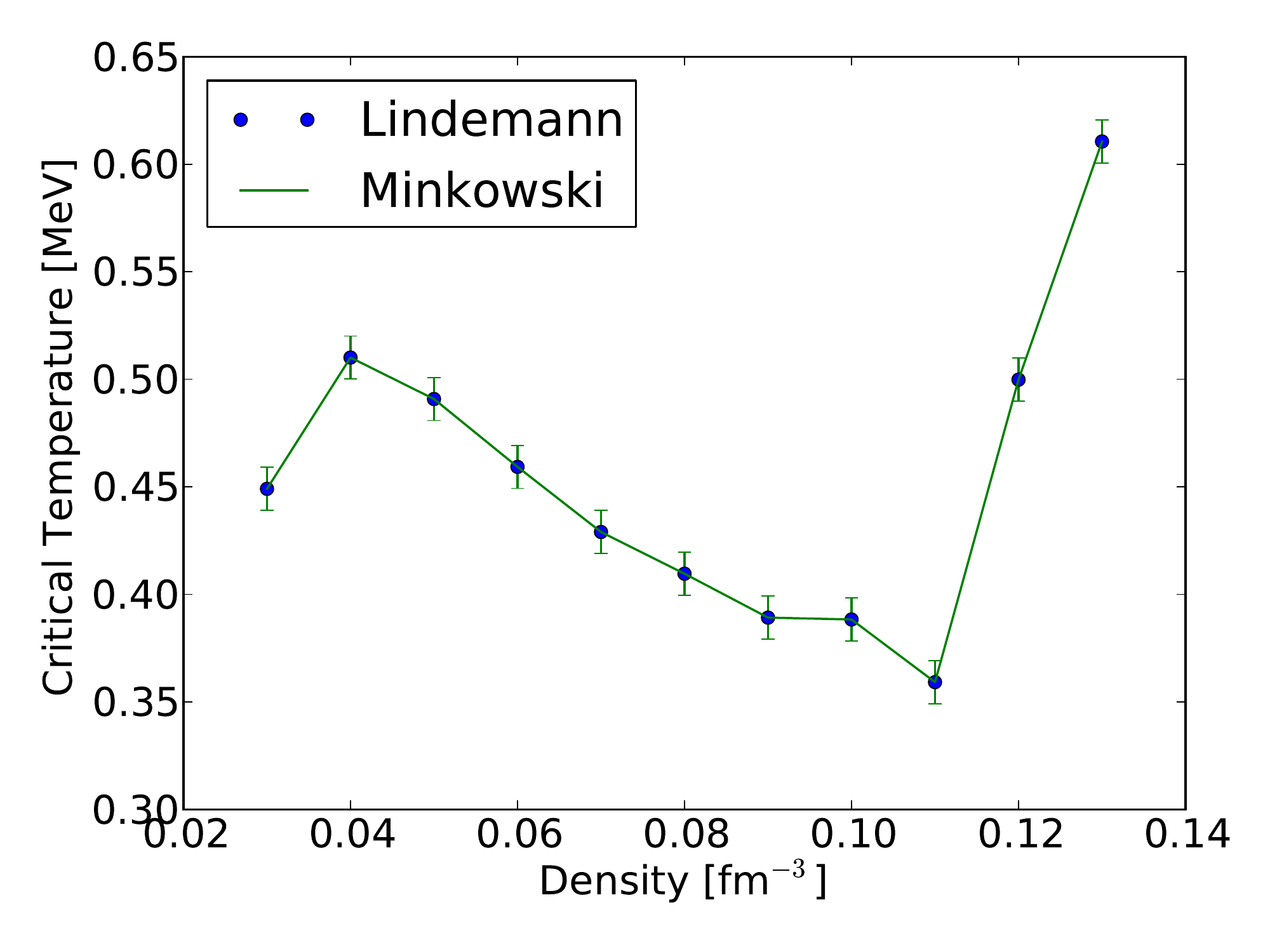}
  \caption{Critical temperature as a function of density. We see the
    overlap between the Minkowski and the Lindeman critical
    temperature.}
  \label{fig:critical_temperature}
\end{figure}

These signals of a solid-liquid phase transition (energy and
Lindemann's coefficient discontinuity) and morphological transition
(Minkowski functionals discontinuity) point at the same transition 
temperature, as can be seen in the phase diagram of
figure~\ref{fig:critical_temperature}.  This means that as the systems
are cooled down at fixed volume, they undergo a thermodinamical and a
morhological phase transition, and they do so at the same temperature.

\subsection{Very Long-range Behavior}
  \label{very_long}
\begin{figure*}[floatfix]  \centering
  \begin{subfigure}[h!]{0.70\columnwidth}
    \includegraphics[width=\columnwidth]{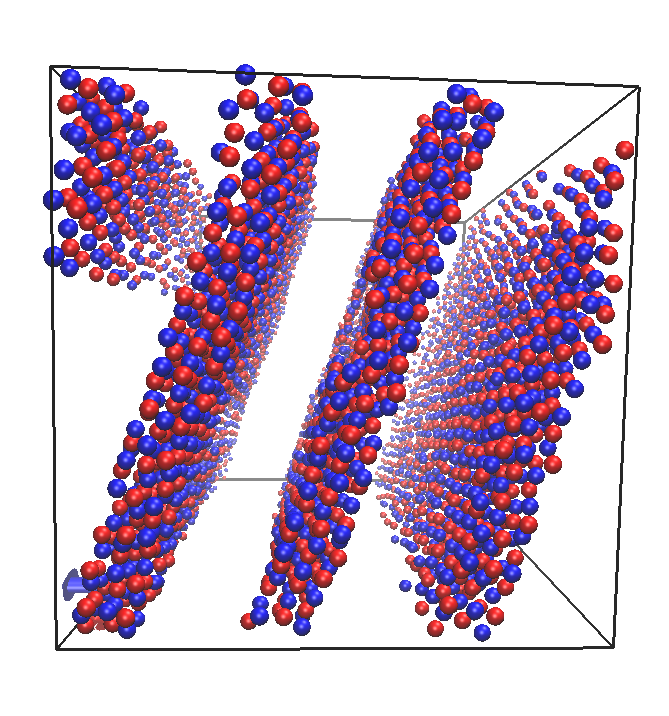}
    \caption*{Below the transition}
  \end{subfigure}
  \begin{subfigure}[h!]{0.70\columnwidth}
    \includegraphics[width=\columnwidth]{morph_0-05_0-50.png}
    \caption*{Above the transition}
  \end{subfigure}
  \caption{Spatial distribution for $\rho=0.05\,\text{fm}^{-3}$, both
    above and below the transition temperature. The structures are
    similar, but much more disordered above the transition.}
  \label{fig:morph}
\end{figure*}

  \makeatletter{}On top of the disappearance of the long range order characteristic of
solids, another feature becomes evident from figure~\ref{fig:rdf}. As
temperature increases through the solid-liquid transition, a very-long
range modulation in the pair correlation function survives. This very
long range ordering is characteristic of the pasta phases. In
figure~\ref{fig:morph}, a visual representation of the spatial
configuration for $\rho=0.05\,\text{fm}^{-3}$ is shown, for
temperatures both below and above the transition. In it, we show that
not only the solid phase has the usual pasta shape, but the liquid
phase preserves it. Below the transition, we have ``frozen
pasta''. Just above it, nucleons may flow but confined to a certain
pasta or pasta-like structure, as we shall see in the following
sections.

\subsection{Neutrino transport properties}
  \label{transport}
  \makeatletter{}This very long-range order, evident from figure~\ref{fig:rdf},
is responsible for a peak at very low momentum $k$ ($\sim 10\,\text{fm}$
wave-length) in the static structure factor $S(k)$, proportional to
the particle scattering probability. With this in mind, we now put the
focus on the very long-range order of our structures.

\begin{figure}
  \includegraphics[width=\columnwidth]{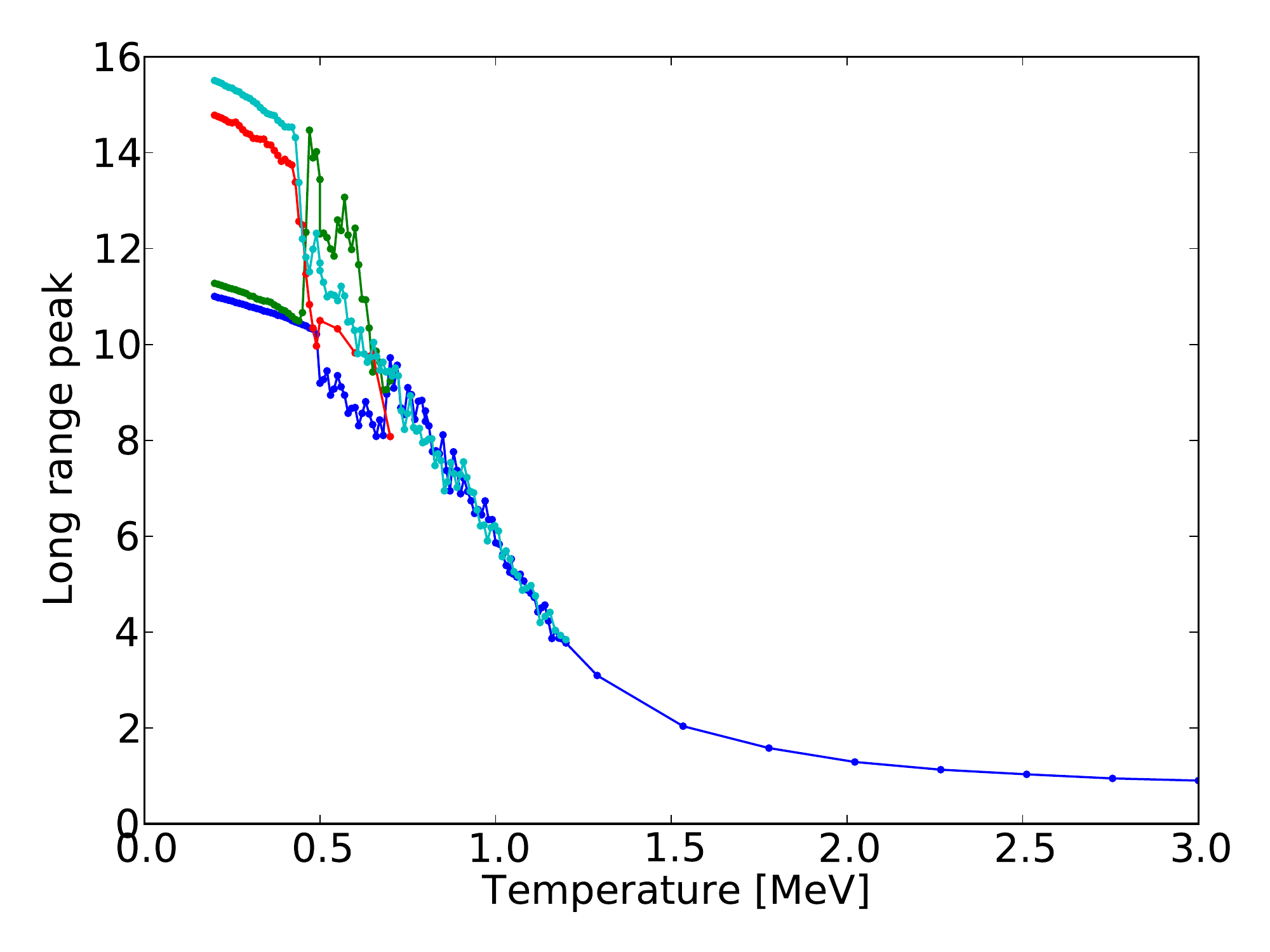}
  \caption{Peak of $S(k)$ for low momenta as a function of the
    temperature, for $\rho=0.05\,\text{fm}^{-3}$. We see two
    behaviors, for $T<0.5\,\text{MeV}$ and $T>0.5\,\text{MeV}$.  }
  \label{fig:sk_peak_0-05}
\end{figure}

In figure~\ref{fig:sk_peak_0-05} we plot the height of the low momenta
peak $S(k<0.5\,\text{fm}^{-1})$ ($\lambda\gtrsim13\,\text{fm}$) as a
function of the temperature for $\rho=0.05\,\text{fm}^{-3}$. The most
clear and intuitive way to read this figure is backwards, from high to
low temperatures, tracing the cooling procedure each system
undergoes in our simulations. Each line in the figure corresponds to
evolutions with different initial conditions but following the same
protocol for cooling down and the same criteria for stability.

At high temperatures the nucleons are rather uniformly distributed 
and no structure is evidenced by $S(k)$: the ``peak'' vanishes as 
its height tends to 1, the value for homogeneous systems. As the
temperature is decreased, a peak at low momentum develops. The
transition described in sections \ref{phase_transition} and 
\ref{morpho}, manifests in figure~\ref{fig:sk_peak_0-05} as 
the vanishing of fluctuations below the transition temperature 
$T \lesssim 0.5\,\text{MeV}$. Even at temperatures as high as 
$T=1.0\,\text{MeV}$ there is still a recognizable low momentum 
absorption peak (with height well over 1), but it does not always 
correspond to a usual pasta (\emph{gnocchi}, \emph{spaghetti} or 
\emph{lasagna}) in our simulations. At such high temperatures and 
for most densities, the system is in a ``sponge-like'' structure 
which is, nevertheless, ordered enough to produce a recognizable 
peak in $S(k)$.

\begin{figure}
  \includegraphics[width=\columnwidth]{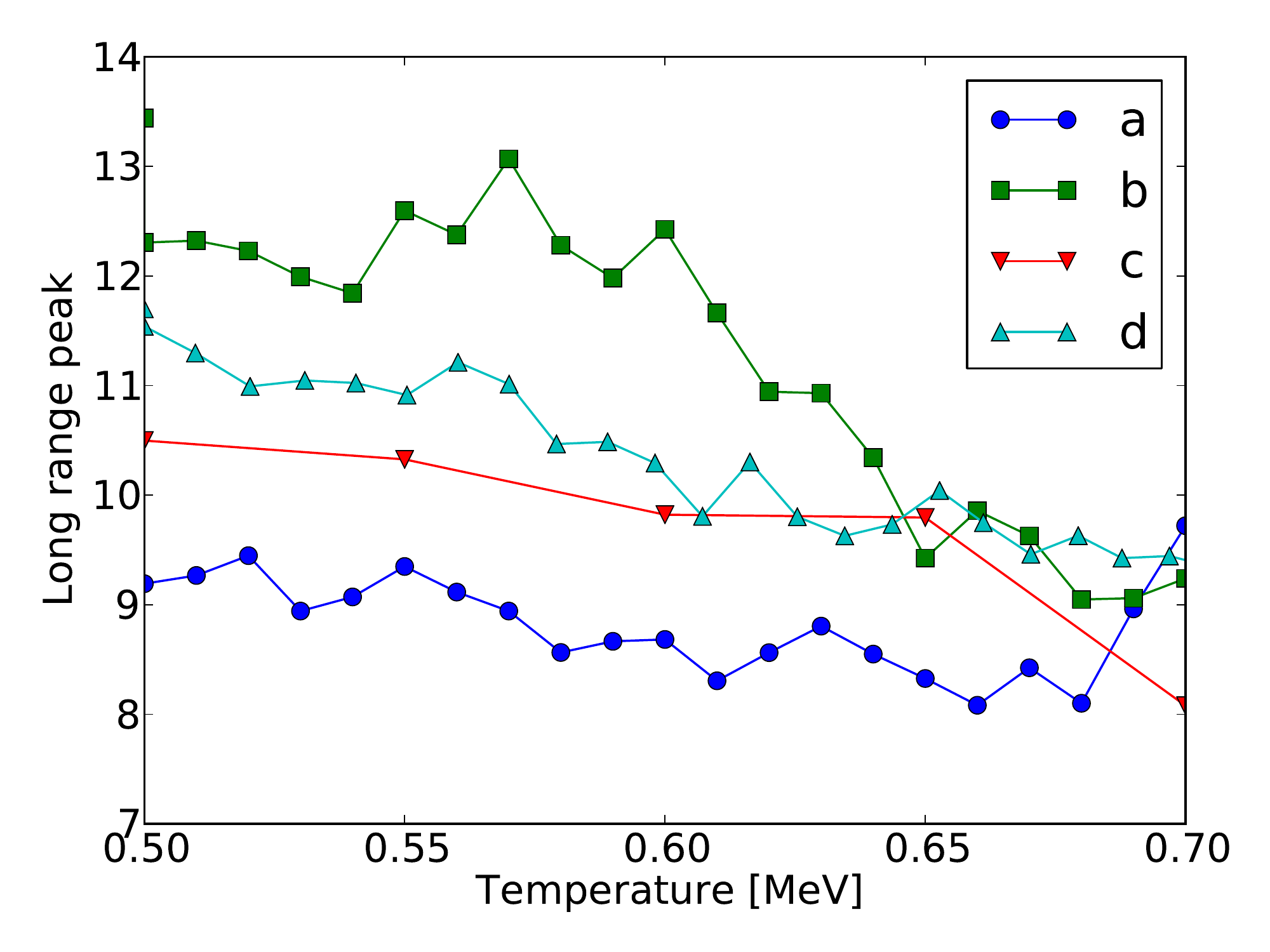}
  \caption{Peak of $S(k)$ for low momenta: zoom into the temperature
    region between $T=0.5\,\text{MeV}$ and $T=0.7\,\text{MeV}$. We can
    see that distinct runs yield different absorption peaks for low
    momenta.}
  \label{fig:sk_peak_zoom}
\end{figure}

\begin{figure*}[floatfix]  \centering
  \begin{subfigure}[h!]{0.7\columnwidth}
    \includegraphics[width=\columnwidth]{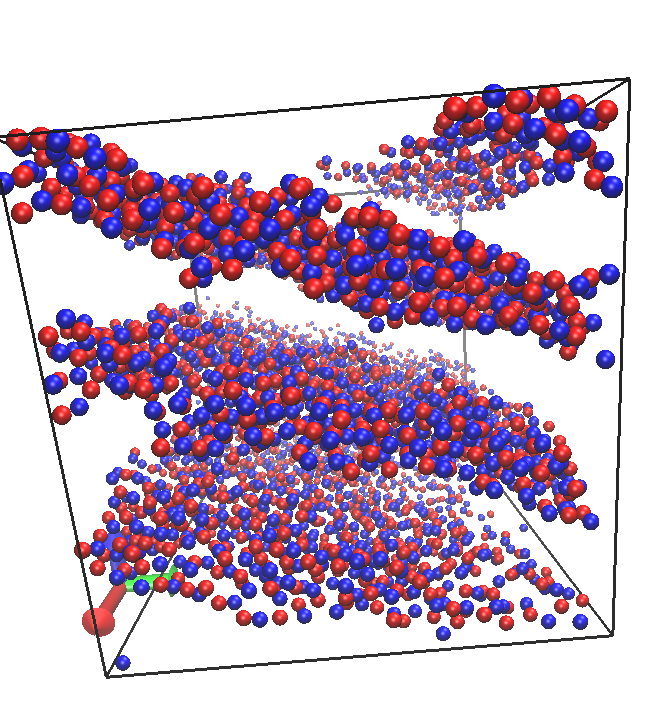}
    \caption{Usual \emph{lasagna}}
  \end{subfigure}
  \begin{subfigure}[h!]{0.7\columnwidth}
    \includegraphics[width=\columnwidth]{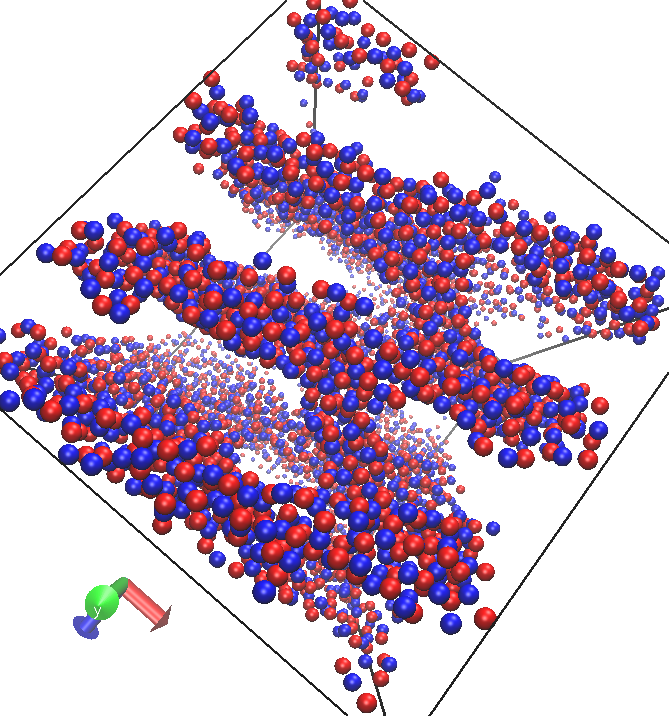}
    \caption{Intertwined \emph{lasagna}}
  \end{subfigure}
  \begin{subfigure}[h!]{0.7\columnwidth}
    \includegraphics[width=\columnwidth]{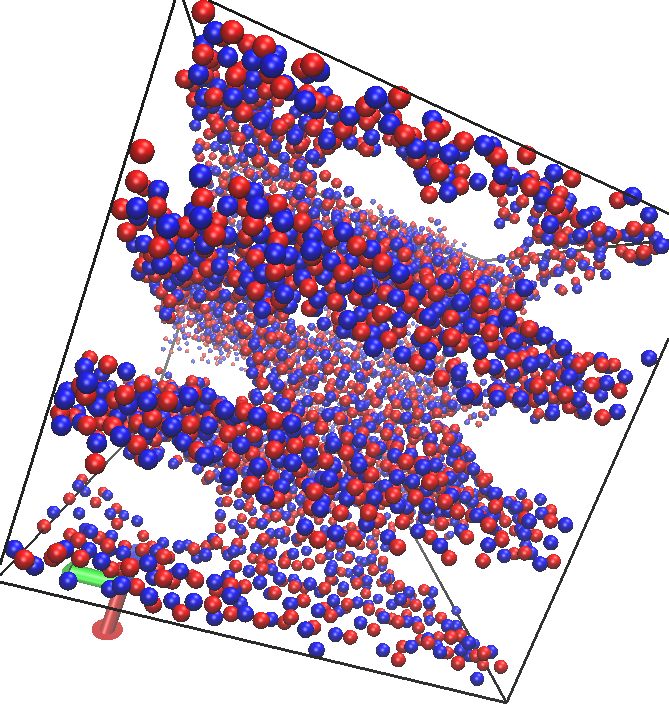}
    \caption{Intertwined \emph{lasagna}}
  \end{subfigure}
  \begin{subfigure}[h!]{0.7\columnwidth}
    \includegraphics[width=\columnwidth]{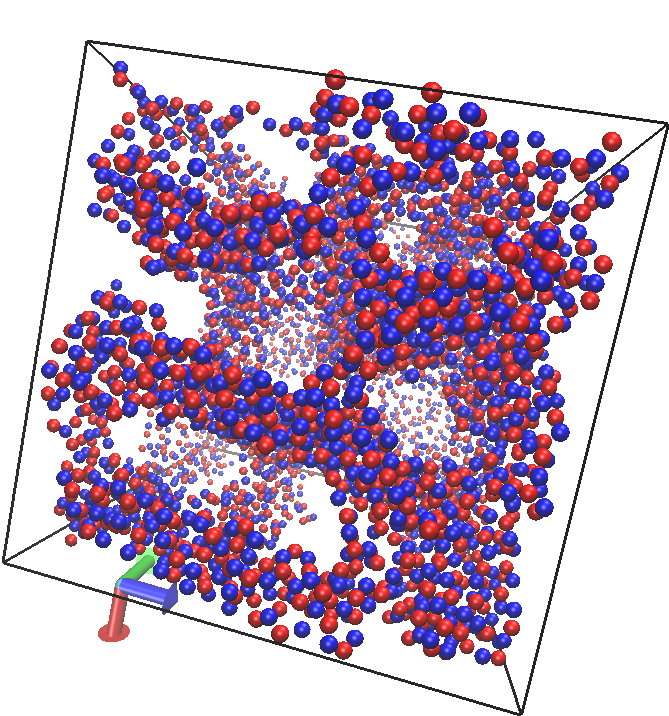}
    \caption{Unusual pasta shape}
  \end{subfigure}
  \caption{Spatial distribution for $\rho=0.05\,\text{fm}^{-3}$ for
    different initial conditions. We see that we have the usual
    \emph{lasagna}, but also intertwined \emph{lasagnas} and other
    structures that don't resemble usual pasta. Despite being
    different from the usual pasta phases, these shapes have a peak
    for low momentum in the structure factor.}
  \label{fig:cool_morph}
\end{figure*}

Interestingly, when the cooling procedure drives the system at
temperatures below $T\sim 0.7\,\text{MeV}$, we observe that in
different runs the system may collapse into several distinct
structures, in addition to the usual lasagna which is the ground state
at this density.  A zoom into the $S(k)$ peak's height for this region
of temperatures can be found in figure~\ref{fig:sk_peak_zoom}, and
snapshots of the structures corresponding to each of those runs can be
seen in figure~\ref{fig:cool_morph} (see captions for details).

\subsection{Properties of non-traditional pasta}
  \label{unusual_pasta}
  \makeatletter{}Usual pasta shapes are ground states (potential energy minima).
The nontraditional structures described in the previous section 
are likely to be local potential energy minima, which abound in 
frustrated systems like this. The complexity of the energy 
landscape (many local minima separated by energy barriers) makes 
it difficult to reach the actual ground state by simple cooling 
in molecular dynamics simulations. However, since we are working 
at fixed number of particles, volume and temperature ($(N,V,T)$ 
ensemble), the equilibrium state of the system at finite 
temperatures is not that which minimizes the internal energy 
but that which minimizes the Helmholtz free energy, 
$A = E - T\,S$. All of these structures may then be actual 
equilibrium solutions, as long as they are free energy minima.

\begin{figure}[floatfix]  \centering
  \begin{subfigure}[h!]{0.95\columnwidth}
    \includegraphics[width=\columnwidth]{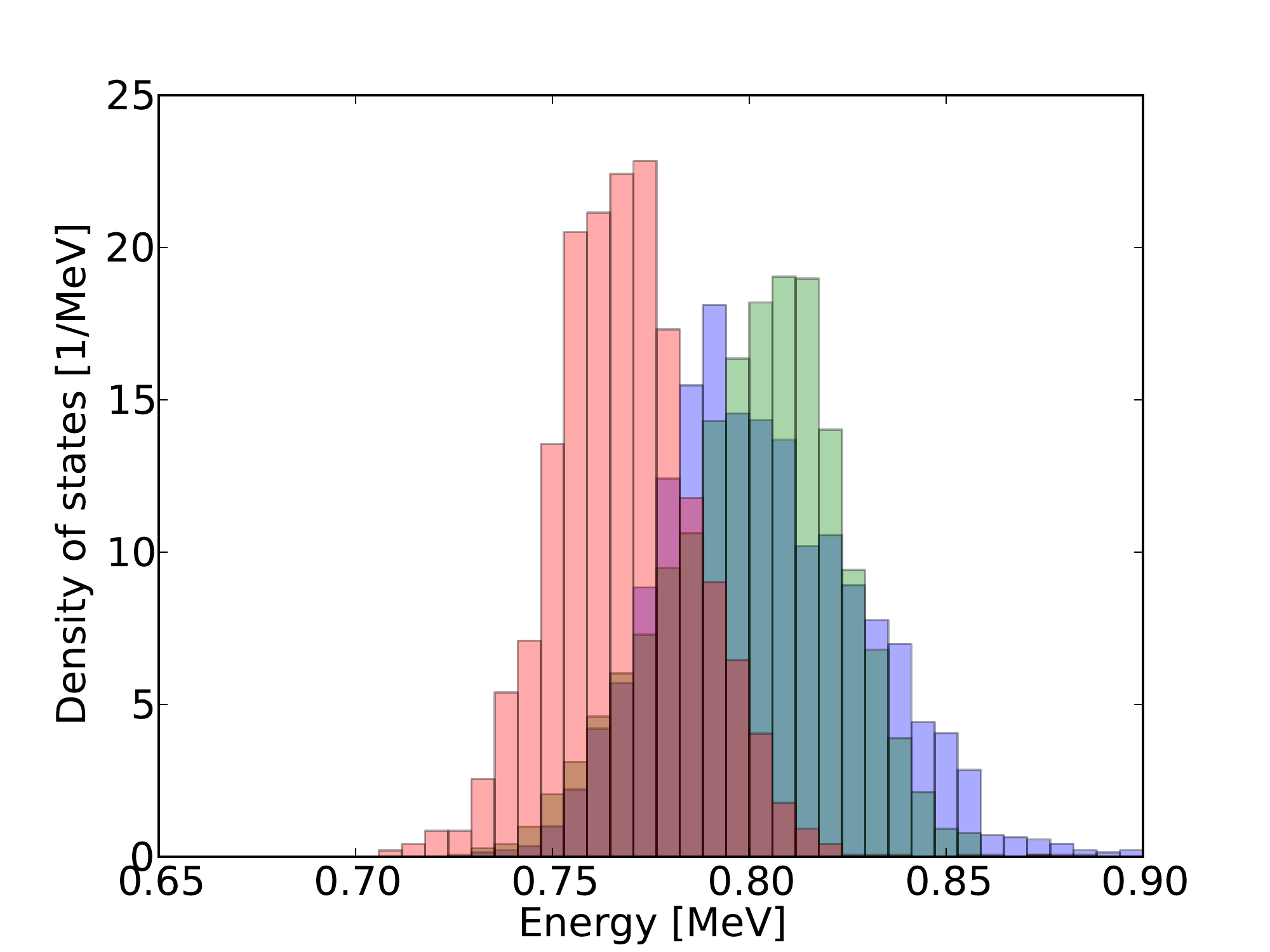}
    \caption{Distribution of energies for $T=0.6\,\text{MeV}$}
    \label{subfig:histo_T_0-06}
  \end{subfigure}
  \begin{subfigure}[h!]{0.95\columnwidth}
    \includegraphics[width=\columnwidth]{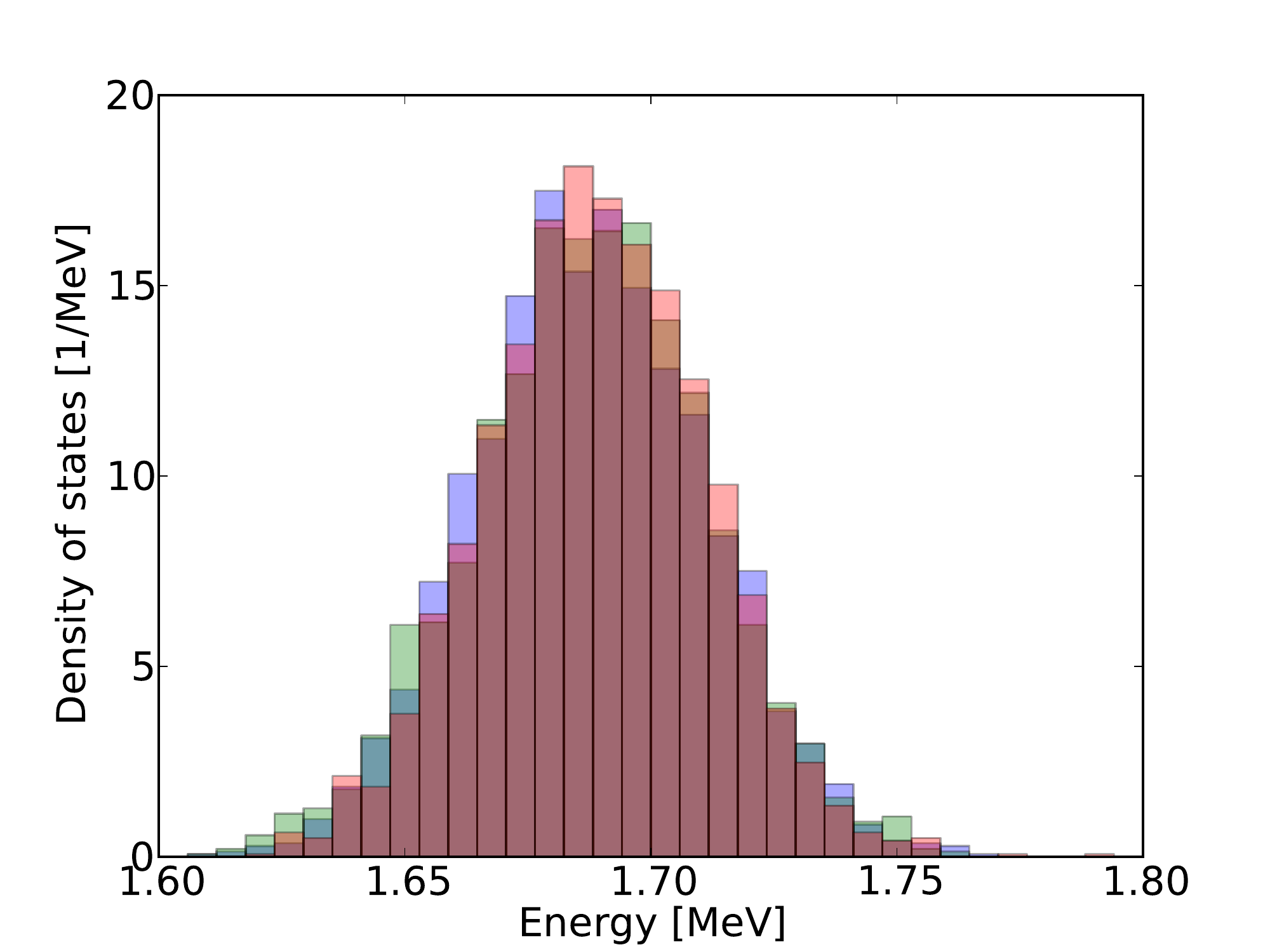}
    \caption{Distribution of energies for $T=0.8\,\text{MeV}$}
    \label{subfig:histo_T_0-08}
  \end{subfigure}
  \caption{Energy distribution for a canonical ensemble. It can 
    be seen that, for $T=0.8\,\text{MeV}$, all three distributions
    overlap completely. However, in $T=0.6\,\text{MeV}$, the histograms, 
    albeit split, still overlap significatively.}
  \label{fig:histo}
\end{figure}

Accurate calculation of free energies from MD simulations is 
computationally very expensive~\cite[pp. 167-200]{frenkel}, 
specially at low temperatures when overcoming energy barriers 
become very improbable events. However, we can easily compute
the internal energy distributions over a long evolution at 
constant temperature. In figure~\ref{fig:histo} we show internal 
energy histograms constructed from very long thermalized 
evolutions at $T=0.6\,\text{MeV}$ using three of the systems 
shown in~\ref{fig:cool_morph} as initial conditions.  We see 
that, although the histograms clearly differ, they overlap 
significantly. This fact indicates that the full ensemble of 
equilibrium configurations at $T=0.6\,\text{MeV}$ contains all 
of these structures, not only lasagna.
In light of this we propose that at low but finite temperatures, 
the state of the system should be described as an ensemble of 
both traditional and nontraditional structures rather than by a 
single one.

When we heat up the system to $T=0.8\,\text{MeV}$, these three
histograms become indistinguishable, hinting that, for this 
temperature, the free energy barriers can be overcome, and 
the system is more likely to be ergodic.

These observations are relevant because all of these structures show 
peaks in $S(k)$ at the same wavelenght (within the uncertainty), 
although of different heights. And more importantly, we find from our 
calculations that the seemingly amorphous, sponge-like structures can be
more efficient in scattering neutrinos of the same momentum that any
usual pasta (i.e. have higher peaks), usually invoked as a necessity
for coherent neutrino scattering. This result shows that unusual pasta
shapes should also be considered when studying the structure of a 
neutron star's crust.

\section{Discussion and concluding remarks}
  \label{discussion}
  \makeatletter{}In this work we used a classical molecular dynamics model to study
symmetric neutron star matter for different temperatures. We found a
solid-liquid phase transition takes place for every density at very 
low temperatures. This transition was characterized with the 
Lindemann coefficient and by a discontinuity in the caloric curves. 
The transition is also signaled by a discontinuity in the Minkowski 
functionals as a function of the temperature, and all three indicators 
gave the same transitions temperature. This phase transition doesn't 
alter the typical pasta shape (\emph{lasagna} and \emph{spaghetti} 
in the cases shown): the liquid phase preserves the pasta shape 
found in the solid phase (the ground state).

As we increase the temperature beyond $T=0.7\,\text{MeV}$, the typical
pasta shapes become unstable and the system adopts slightly less 
ordered but still inhomogeneous structures. However, the low momentum 
absorption peak of this structures remains quite high. This implies 
that the existence of traditional pasta shapes --which are only 
obtained at extremely low temperatures-- is not a necessary condition 
for the enhancement of the neutrino absorption in a neutron star's crust. 

Furthermore, we also found that at $T\sim0.7,\text{MeV}$ the system 
can exist in various stable states, all of them with different 
morphology --and, consequently, different structure factor--, but 
very close in internal energy. From our simulations at fixed 
$(N,V,T)$, these states appear to be separated by relatively high 
energy barriers which make the spontaneous transition between them 
a very improbable event, and unlikely to be observed within a single 
simulation run. However, the energy distributions obtained from 
long enough runs starting from different states overlap significantly, 
indicating that all of them are members of the full ensemble of 
equilibrium states at that temperature.
At $T\sim0.8\,\text{MeV}$ the energy barriers become surmountable and 
the energy histograms completely overlap.

All of this suggests that the actual state of these systems at low, but 
finite temperatures, is better described as an ensemble of shapes rather 
than by a single pasta-like structure.

\begin{acknowledgments}
  C.O.D. is a member of the Carrera de Investigador CONICET, work
  partially supported by CONICET Grant PIP0871. P.G.M. and P.N.A.  by
  a CONICET grant. The three-dimensional figures were prepared using
  Visual Molecular Dynamics~\cite{VMD}.
\end{acknowledgments}

\end{document}